\documentclass[preprint,aps,floats,subeqn,showpacs]{revtex4}
\usepackage{epsfig}
\newcommand{\be}{\begin{equation}}
\newcommand{\ee}{\end{equation}}
\newcommand{\bea}{\begin{eqnarray}}
\newcommand{\eea}{\end{eqnarray}}
\newcommand{\bml}{\begin{mathletters}}
\newcommand{\eml}{\end{mathletters}}

\begin{document}

\tighten

\preprint{IUB-TH-043}
\draft

%\twocolumn[\hsize\textwidth\columnwidth\hsize\csname @twocolumnfalse\endcsname

%%%%%%%%%%%%%%%%%%%%%%%%%%%%%%%%%%%%%%%%%%%%%%%%%%%%%%%%%%%%%%%%%%%%%%%%%%

%\wideabs{    % Uncomment this line for two-column output

\title{Spherically symmetric solutions of
a $(4+n)$- dimensional Einstein-Yang-Mills model with cosmological constant}
\renewcommand{\thefootnote}{\fnsymbol{footnote}}
\author{ Yves Brihaye\footnote{Yves.Brihaye@umh.ac.be}}
\affiliation{Facult\'e des Sciences, Universit\'e de Mons-Hainaut,
7000 Mons, Belgium}
\author{Betti Hartmann\footnote{b.hartmann@iu-bremen.de}}
\affiliation{School of Engineering and Sciences, International University Bremen (IUB),
28725 Bremen, Germany}

\date{\today}
\setlength{\footnotesep}{0.5\footnotesep}

%%%%%%%%%%%%%%%%%%%%%%%%%%%%%%%%%%%%%%%%%%%%%%%%%%%%%%%%%%%%%%%%%%%%%%%%%%
\begin{abstract}
We construct solutions of an Einstein-Yang-Mills
system including a cosmological constant in $4+n$ space-time dimensions, where
the $n$-dimensional manifold associated with
the extra dimensions is taken to be Ricci flat.
Assuming the matter and metric fields to be 
independent of the $n$
extra coordinates, a spherical symmetric Ansatz for
the fields leads to a set of
coupled ordinary differential equations. 
We find that for $n > 1$ only solutions with either one non-zero Higgs field
or with all Higgs fields constant and zero gauge field function (corresponding
to a Wu-Yang-type ansatz) exist. 
We give the analytic solutions available
in this model. These are ``embedded'' abelian solutions with a diverging size of the
manifold associated with the extra $n$ dimensions. Depending on the choice
of parameters, these latter solutions either represent naked singularities
or they possess a single horizon.

We also present solutions of the effective 4-dimensional
Einstein-Yang-Mills-Higgs-dilaton model, where the higher dimensional
cosmological constant induces a Liouville-type potential.
The solutions
are non-abelian solutions with diverging Higgs fields, which
exist only up to a maximal value of the cosmological constant.

\end{abstract}

\pacs{04.20.Jb, 04.40.Nr, 04.50.+h, 11.10.Kk }
\maketitle

\section{Introduction}
In an attempt to unify electrodynamics and general relativity,
Kaluza introduced an extra, a fifth dimension \cite{kaluza} and assumed all
fields to be independent of the extra dimension.
Klein \cite{klein} followed this idea, however, he assumed the fifth dimension
to be compactified on a circle of Planck length. The resulting theory
describes $4$-dimensional Einstein gravity plus Maxwell's equations. 
One of the new fields appearing in this model
is the dilaton, a scalar companion of the metric tensor.
In an analogue way, this field arises in the
low energy effective action of superstring theories and is
associated with the classical scale invariance of these models \cite{maeda}.

Both string theories \cite{pol} as well as so-called ``brane worlds'' 
\cite{brane} assume that space-time possesses more than four dimensions.
In string theories, these extra dimensions are -following the idea of Klein-
compactified on a scale of the Planck length, while in brane worlds,
which assume the Standard model fields to be confined on a 3-brane, they are
large or even infinite. It should thus be interesting to study classical solutions
of non-abelian gauge theories in higher dimensions and with view
to the AdS/CFT \cite{ads}, respectively dS/CFT \cite{ds} correspondence especially
including a cosmological constant. 

Classical soliton-like solutions of Einstein-Yang-Mills theories in
$d$ dimensions where studied in \cite{bcht}. 
These solutions are spherically symmetric in the full $d$ dimensions.
Solutions with spherical symmetry in only $3+1$ dimensions
were considered for $n=1$ codimension in \cite{volkov}, while they
were studied for $n$ codimensions in \cite{bch}. 
The manifold associated with the extra dimensions was assumed to be
Ricci flat. If all fields are assumed to be independent of
the extra coordinates, the $4+n$ dimensional
system reduces to an effective $4$-dimensional 
Einstein-Yang-Mills-Higgs-dilaton (EYMHD) system with $n$ independent, massless 
dilatons and $n$ Higgs triplets. Constraints appear in this model
such that only solutions with either one non-zero Higgs field
or with all Higgs fields constant and zero gauge
field function (Wu-Yang-Type ansatz) exist.

If a cosmological constant is introduced into this model, 
the effective $4$-dimensional
action describes an EYMHD system, where now the dilatons become
massive due to a Liouville-type potential. The $(4+1)$-dimensional
version of this model was studied in \cite{bbh1}. It was found that
no asymptotically flat nor de Sitter nor Anti-de Sitter solutions
exist in this model- analogue to the Einstein-Maxwell-dilaton
case with Liouville potential \cite{pw}. Consequently, black hole solutions with unusual asymptotics
were constructed, which manifest the ``embedded'' abelian counterparts
of the $4$-dimensional solutions found in \cite{chm}.

In this paper, we study the system of \cite{bbh1} in 
$4+n$ dimensions, assuming all
fields to be independent of the $n$ extra coordinates.
In section II, we present the model including the Ansatz, 
the equations and 
the analytic solutions, which fufill the constraints that arise in this
model.
In Section III, we give the
$4$-dimensional effective action resulting
from the model and give analytic solutions, which represent
non-abelian solutions with diverging Higgs fields. These solutions 
exist only up to a maximal value of the cosmological constant.
We give our conclusions in Section IV.

%%%%%%%%%%%%%%%%%%%%%%%%%%%%%%%%%%%%%%%%%%%%%%%%%%%%%%%%%%%%
\section{The $(4+n)$-dimensional $\Lambda$-Einstein-Yang-Mills model}
%%%%%%%%%%%%%%%%%%%%%%%%%%%%%%%%%%%%%%%%%%%%%%%%%%%%%%%%%%%%

The $\Lambda$-Einstein-Yang-Mills Lagrangian
in $d=4+n$ dimensions is
given by:

\begin{equation}
\label{action}
  S = \int \Biggl(
    \frac{1}{16 \pi G_{(4+n)}}\left( R - 2 \Lambda_{(n+4)} \right)  - 
    \frac{1}{4 e^2}F^a_{M N}F^{a M N}
  \Biggr) \sqrt{g^{(4+n)}} d^{(4+n)} x
\end{equation}
with the SU(2) Yang-Mills field strengths
$F^a_{M N} = \partial_M \mathsf{A}^a_N -
 \partial_N \mathsf{A}^a_M + \epsilon_{a b c}  \mathsf{A}^b_M \mathsf{A}^c_N$
, the gauge index
 $a=1,2,3$  and the space-time index
 $M=0,...,(4+n)-1$. $G_{(4+n)}$ and $e$ denote
respectively the $(4+n)$-dimensional Newton's constant and the coupling
constant of the gauge field theory. $G_{(4+n)}$ is related to the Planck mass
$M_{pl}$ by $G_{(4+n)}=M_{pl}^{-(2+n)}$ and $e^2$ has the dimension of $[{\rm length}]^n$.
$ \Lambda_{(n+4)}$ is the $(4+n)$-dimensional cosmological constant.

In the following, we denote the coordinates $x_{(3+k)}$ by $y_k$ with 
$k=1,...,n$.

If both the matter functions and the metric functions
are independent on $y_k$, the fields can be
parametrized as follows:
\begin{equation}
\label{mansatz}
g^{(4+n)}_{MN}dx^M dx^N = 
e^{-\Xi}g^{(4)}_{\mu\nu}dx^{\mu}dx^{\nu}
+\sum_{k=1}^n e^{2\zeta_k} (dy^k)^2 
\ , \ \mu , \nu=0, 1, 2, 3 
\end{equation}
with
\begin{equation}
\Xi=\sum_{k=1}^n \zeta_k 
\end{equation}
and
\begin{equation}
\mathsf{A}_M^{a}dx^M=\mathsf{A}_{\mu}^a dx^{\mu}+ \sum_{k=1}^n \Phi_k^a dy^k   \  . \
\end{equation}
$g^{(4)}$ is the $4$-dimensional metric tensor 
and the $\zeta_{j}$ and  $\Phi_j^a$, $j=1,...,n$,  
play the
role of  dilatons and Higgs fields, respectively. 
Note that the factor $e^{2\zeta_k}$ leads to a space-time dependent
seize of the manifold associated with the extra dimensions.
The same holds true for the factor $e^{-\Xi}$ with respect to the $4$-dimensional
``physical'' manifold. 
 
The case $n=1$, $\Lambda_{5}=0$ was studied in \cite{volkov}, while
the case $n=1$, $\Lambda_{5}\neq 0$ was investigated in \cite{bbh1}.
This present paper is an extension of the results in \cite{bch}, which deals
with the case $\Lambda_{n+4}=0$ for generic $n$.

%%%%%%%%%%%%%%%%%%%%%%%%%%%%%%%%%%%%%%%%%%%%
\subsection{Spherically symmetric Ansatz}
%%%%%%%%%%%%%%%%%%%%%%%%%%%%%%%%%%%%%%%%%%%%%%

For the metric the spherically symmetric Ansatz
in Schwarzschild-like coordinates reads \cite{weinberg}:
\begin{equation}
g^{(4)}_{\mu\nu}dx^{\mu}dx^{\nu}=
-A^{2}(r)N(r)dt^2+N^{-1}(r)dr^2+r^2 d\theta^2+r^2\sin^2\theta
d^2\varphi
\label{metric}
\ , \end{equation}
with
\begin{equation}
N(r)=1-\frac{2m(r)}{r}
\ . \end{equation}

For the gauge and Higgs fields, we use the purely magnetic hedgehog ansatz
\cite{thooft} :
\begin{equation}
{\mathsf{A}_r}^a={\mathsf{A}_t}^a=0
\ , \end{equation}
\begin{equation}
{\mathsf{A}_{\theta}}^a= (1-K(r)) {e_{\varphi}}^a
\ , \ \ \ \
{\mathsf{A}_{\varphi}}^a=- (1-K(r))\sin\theta {e_{\theta}}^a
\ , \end{equation}
\begin{equation}
\label{higgsansatz}
{\Phi}^a_{j}=c_j v H_j(r) {e_r}^a \ \ , \ \ j=1,...,n \ ,
\end{equation}
where $v$ is a mass scale, while $c_j$ are dimensionless constants
determining the vacuum expectation values of the Higgs fields 
$\langle \Phi_j \rangle = c_j v$.
In absence of a Higgs potential these have to be set by hand.
Finally, the dilatons are scalar fields depending only on $r$~:
\begin{equation}
\zeta_j =\zeta_j(r) \ \ , \ \ j = 1,...,n 
\ . \end{equation}

%%%%%%%%%%%%%%%%%%%%%%%%%%%%%%%%%
\subsection{Equations of motion}
%%%%%%%%%%%%%%%%%%%%%%%%%%%%%%%%%%
The non-vanishing components of the energy-momentum tensor are given by:
\begin{eqnarray}
& T_0^0 &= - e^{\Xi}
\left(\sum_{k=1}^n (\mathcal{A}_k  + C_k ) + B + D\right) \ , \nonumber \\
& T_1^1 &= - e^{\Xi}
\left(\sum_{k=1}^n (-\mathcal{A}_k  + C_k ) - B + D\right) \ , \nonumber \\
& T_2^2 &= T_3^3 = - e^{\Xi}
\left(\sum_{k=1}^n \mathcal{A}_k  - D\right) \ ,  \nonumber \\
& T_j^j &= - e^{\Xi}
\left(-2 (\mathcal{A}_j+C_j) 
+ \sum_{k=1}^n (\mathcal{A}_k + C_k) + B + D\right) \ , \ j=1,...,n \ , \nonumber \\
& T_j^k & = e^{\Xi} e^{-(\xi_1+\xi_2)}\left(\frac{N}{2} H'_j H'_k + \frac{K^2}{x^2} H_j H_k\right) \ , \ j\neq k \ ,
\end{eqnarray}
 where we use the abbreviations
 \be
 \mathcal{A}_j = \frac{1}{2} e^{-2 \zeta_j} N (H_j')^2 \ \ , \ \
 C_j = e^{-2 \zeta_j} K^2 H_j^2 \frac{1}{x^2} \ \  \ \
 \ee
 and
 \be
  B =  e^{\Xi} \frac{1}{x^2} N (K')^2 \ \ , \ \
  D =  e^{\Xi} \frac{1}{2 x^4}  (K^2 - 1)^2  \ .
 \ee

Defining the coupling constant $\alpha = v \sqrt{G_{(4+n)}}$, the rescaled
cosmological constant $\Lambda= \frac{1}{e^2 v^2} \Lambda_{(4+n)}$, the 
mass function $\mu=evm$ and the radial coordinate $x= e v r$,
we obtain the following differential equations (the prime denotes the derivative
with respect to $x$) :
\be
\label{mu}
\mu' = \alpha^2 x^2\left(\sum_{k=1}^n (\mathcal{A}_k+C_k) 
+ B + D\right) + \frac{1}{2}N x^2 \Theta  + \frac{\Lambda x^2}{4} e^{-\Xi}
\ee
\be
\label{A}
   A' = \alpha^2 A x \left(\sum_{k=1}^n e^{-2\zeta_k} (H_k')^2 
   + 2 e^{\Xi} \frac{K'^2}{x^2}\right) + x A \Theta
\ee
with 
\be
  \Theta \equiv \frac{3}{4}\left(\sum_{k=1}^n (\zeta_k')^2 
+ \frac{2}{3} \sum_{k > k'} \zeta_{k}' \zeta_{k'}' \right)  \ ,
\ee
\begin{equation}
\label{zeta}
(x^2 A N \zeta_j ')'
= \alpha^2 A x^2 e^{-\Xi}
\left
[ \frac{2}{n+2}\sum_{M=0}^
{d-1} T_M^M -2 T_j^j\right] - 
\frac{A x^2}{n+2}\Lambda e^{-\Xi} \  \ , \ \ j=1,...,n \ ,
\end{equation}
\be
\label{k}
   (e^{\Xi}ANK')'
   = A\left( \frac{1}{x^2}e^{\Xi}K(K^2-1)
       + \sum_{k=1}^n e^{-2 \zeta_k} K H_k^2\right)  \ ,
\ee
\be
\label{h}
   (e^{-2\zeta_j}x^2 A N H_j')' = 2 A e^{-2 \zeta_j} K^2 H_j \ \ , \ \
   j=1,...,n   \ .
\ee
Finally, since the off-diagonal components of the Einstein tensor vanish,
we obtain constraints on the fields from the $jk$-components of the energy-momentum
tensor:
\begin{equation}
\label{constraint}
 e^{\Xi} e^{-(\xi_1+\xi_2)}\left(\frac{N}{2} H'_j H'_k 
+ \frac{K^2}{x^2} H_j H_k\right)=0 \ , \ j\neq k \ .
\end{equation}
These constraints lead to the conclusion that solutions
with either only one non-zero Higgs field
or with all Higgs fields constant and zero gauge field function $K$
exist. The former case would effectively correspond to the $4+1$-dimensional
model with one Higgs field. In the following, we will discuss the solutions
available in $4+n$ dimensions with all Higgs fields being constant. 

%%%%%%%%%%%%%%%%%%%%%%%%%%%%%%%%%%%%%%%%%%%%
\subsection{ ``Embedded'' abelian solutions}
%%%%%%%%%%%%%%%%%%%%%%%%%%%%%%%%%%%%%%%%%%%%
We assume here that  $H_1(x)=H_2(x)=...\equiv H(x)$ and thus
$\zeta_1(x)=\zeta_2(x)=...\equiv \zeta(x)$.

Note that no solutions with constant seize of the
manifold associated with the extra dimensions (i.e. constant dilaton fields)
exist in this model, if one chooses either $K(x)=0$, $H_k(x)=1$ 
or $K(x)=1$, $H_k(x)=0$. However, solutions with non-constant dilatons
exist.
Choosing
\begin{equation}
K(x)=0 \ \ \ , \ \ \ H(x)=1 \ \ ,
\end{equation}
the solutions correspond to ``embedded'' abelian solutions with magnetic charge
$P=1$ since $F^{(r)}_{\theta\varphi} = \sin\theta$. 

The counterparts of these solutions  
for a model involving one dilaton field with Liouville type potential
in 4-dimensional Einstein-Maxwell theory were discussed in \cite{chm}. 
Note that the effective 4-dimensional
action which results from our model
in the case $n=1$ \cite{bbh1}  
corresponds to the $4$-dimensional model studied in \cite{chm}.
Here, however, the effective action contains $n$ dilaton fiels.
As was proven in \cite{pw}, no asymptotically flat nor de Sitter nor Anti-de Sitter
solutions exist for the model in \cite{chm}. This is also true
for the ``embedded'' abelian counterparts \cite{bbh1} and 
the system studied here.

Here we present the $(4+n)$-dimensional counterparts of
the solutions found in \cite{bbh1}. The solutions read:
\begin{equation}
\label{ansatzemd}
A(x)=a_0 x^{(n+2)/2} \ \ , \ \ N(x)=n_0-n_1 x^{-2(n+1)/n} \ \ , \ \ \zeta(x)=\zeta_0 + \frac{2}{n} \ln(x)
\end{equation}  
where $a_0$, $n_1$ and $\zeta_0$ are not further determined constants, while 
\begin{equation}
n_0=\frac{n^2 \alpha^2}{2(n+2)(n+1)} e^{n\zeta_0}-\frac{\Lambda n^2}{4(n+1)(n+2)} e^{-n\zeta_0}
\end{equation} 
and the cosmological constant is given by:
\begin{equation}
\Lambda=(n+2) e^{n\zeta_0} - 2(n+1) \alpha^2 e^{2n\zeta_0}  \ .
\end{equation}
The metric then reads:
\begin{eqnarray}
ds^2&=&-e^{-n\zeta_0} a_0^2 x^{n} \left(n_0-n_1x^{-(2n+2)/n}\right) dt^2+
e^{-n\zeta_0} x^{-2} \left(n_0-n_1 x^{-(2n+2)/n}\right)^{-1} dx^2 \nonumber \\ 
&+&
e^{-n\zeta_0} \left(d\theta^2 +\sin^2\theta d\varphi^2\right) 
+ e^{2\zeta_0} x^{4/n} \left( (dy_1)^2 + ...+ (dy_n)^2\right)
\end{eqnarray}

The Kretschmann
scalar $K=R^{MNOP} R_{MNOP}$ of the solutions is in general
very compicated, that's why we give its form here for $n=2$. It reads:
\begin{equation}     
K_{(n=2)} =\frac{2 e^{4\zeta_0}}{3}\left(
\alpha^4( 9 e^{12\zeta_0}  + 6 e^{8\zeta_0}+ e^{4\zeta_0} 
) - \alpha^2(12 e^{10\zeta_0} - 4 e^{8\zeta_0})  
 + 4 e^{8\zeta_0} +
6  + 18\frac {n_1^2}{x^6}\right)
\end{equation}
Clearly, the solutions possess a physical singularity at $x=0$.
This is also true for $n\neq 2$, e.g. for $n=1$ and $n=3$ the leading 
divergence is proportional to $x^{-10}$, respectively $x^{-\frac{10}{3}}$.
This clearly indicates that the singularity gets weaker when increasing
the number of extra dimensions. We also note that for $n\neq 2$, the
Ricci scalar $R$ and the invariant $R^{AB} R_{AB}$ are singular at
$x=0$, e.g. for $n=1$ the leading divergence is of power $x^{-5}$ and 
$x^{-10}$ for $R$, respectively $R^{AB} R_{AB}$. However, for $n=2$
$R$ and $R^{AB} R_{AB}$ are constant:
\begin{equation}
R_{(n=2)}=
2e^{2\zeta_0}\left(3 e^{6\zeta_0}\alpha^2-
2e^{4\zeta_0}+e^{2\zeta_0}\alpha^2-1\right)  \ ,
\end{equation}
\begin{equation}
(R^{AB}R_{AB})_{(n=2)}=e^{4\zeta_0}\left(9e^{12\zeta_0} 
\alpha^4-12 e^{10\zeta_0}\alpha^2+
6e^{8\zeta_0}\alpha^4+4e^{8\zeta_0}-4e^{6\zeta_0}\alpha^2+e
^{4\zeta_0}\alpha^4+2\right)
\end{equation}
This can be related to the particular ansatz that we have introduced for the 
metric (see (\ref{mansatz})). Clearly for $n=2$, the prefactor in front
of the 4-dimensional, spherically symmetric part becomes $e^{-2\zeta}$, while
the prefactor in front of the extra dimensional part is always $e^{2\zeta}$.
For $n=2$ we thus have $g_{tt}\propto g_{xx}$.  

Coming now back to the interpretation of the solutions, we note that 
excluding the choice of $n_0 < 0$ and $n_1 > 0$ (which would correspond
to a space-like naked singularity), this solution represents
a naked singularity for $n_0 > 0$, $n_1 < 0$, while
a horizon exists at $x_h=\left(\frac{n_1}{n_0}\right)^{n/(2n+2)}$
for a) $n_0 > 0$, $n_1 > 0$ and  b) $n_0 < 0$, $n_1 < 0$.
For case a), $n_0$ is always positive for $\Lambda < 0$, but
also for $0 < \Lambda < 2\alpha^2 e^{2n\zeta_0}$. For b) we have to choose
$\Lambda$ positive with $\Lambda > 2\alpha^2 e^{2n\zeta_0}$.
Moreover, the seize associated with the extra dimension- manifold diverges
at infinity. A similar behaviour of the solutions was observed 
in a model without gauge fields \cite{wiltshire}. The solutions
thus correspond to black holes (if $n_0$ and $n_1$ are chosen
appropriately) with toroidal horizon (note the constant seize of
the 2-spheres) which is extended non-trivially into $n$ extra dimensions.

The temperature $T$ 
of static black hole solutions is given by:
\begin{equation}
\tilde{T}=
\frac{1}{2\pi}\sqrt{-\frac{1}{4}g^{tt} g^{MN}\left
(\partial_M g_{tt} \right)\left(\partial_N g_{tt}\right)}  \ .
\end{equation}
Evaluation at the horizon $x_h$ then gives for the black hole solutions studied
here (with $T=\frac{\tilde{T}}{ev}$):
\begin{equation}
T=\frac{a_0 n_1 (n+1)}{2\pi n} x_h^{\frac{n}{2}-\frac{2}{n}-2} \ .
\end{equation}
As can be easily checked, the derivative of $T$ with respect to $x_h$,
$\frac{\partial T}{\partial x_h}$, changes sign
for $n=-1$ and $n=2\pm 2\sqrt{2}$. Since we are only interested
in positive $n$ here, we find that $\frac{\partial T}{\partial x_h} < 0$
for $n \le 4$ and  $\frac{\partial T}{\partial x_h} > 0$
for $n \ge 5$. 

\begin{figure}
\centering
\epsfysize=20cm
\mbox{\epsffile{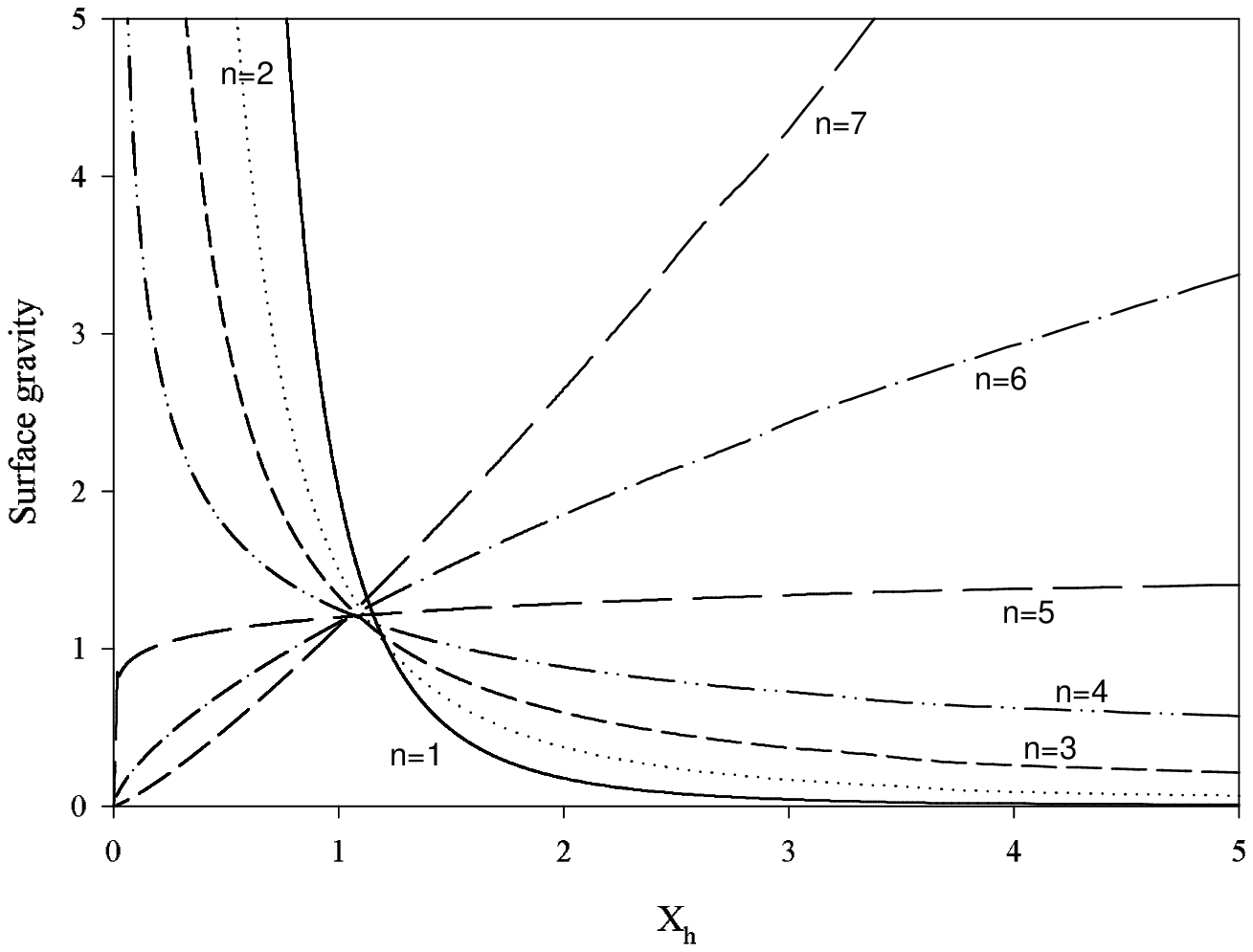}}
\caption{\label{figkappa} The surface gravity of the embedded
abelian black hole solutions, which is equal to
$2\pi T$, is shown in
dependence on the horizon value $x_h$ for $n=1,...,7$.}
\end{figure}

For the choice of the parameters $a_0=1$, $\zeta_0=0$ and $n_1=1$, 
we plot $2\pi T$ (which is actually equal to
the surface gravity) over $x_h$ for different values of $n$ in Fig. \ref{figkappa}.

%%%%%%%%%%%%%%%%%%%%%%%%%%%%%%%%%%%%%%%%%%%%%
\section{An effective 4-dimensional model}
%%%%%%%%%%%%%%%%%%%%%%%%%%%%%%%%%%%%%%%%%%%%%%

 As in the 5-dimensional case \cite{volkov} (in our notation $n=1$) the 
equations given in the previous section can equally well be derived 
from an effective 4-dimensional
Einstein-Yang-Mills-Higgs-dilaton (EYMHD) Lagrangian.
In this section, we will discuss the following matter Lagrangian:
\begin{eqnarray}
\label{effeL}
   L_M =
&-& \frac{1}{4} e^{2 \omega \Gamma }F^a_{\mu \nu} F^{a, \mu \nu}
- \sum_{k=1}^n 
\frac{1}{2} e^{-4 \omega \Psi_k} D_{\mu} \Phi_k^a D^{\mu} \Phi_k^a
\nonumber \\
&-& \frac{1}{2} 
\left(\sum_{k=1}^n \partial_{\mu} \Psi_k  \partial^{\mu} \Psi_k
        +\frac{2}{3} \sum_{k > k'} \partial_{\mu} \Psi_k  \partial^{\mu} 
\Psi_{k'}\right)
- \frac{\tilde \Lambda}{2} e^{-2 \omega \Gamma}
\end{eqnarray}
with
\begin{equation}
\Gamma=\sum_{k=1}^n \Psi_k  \ .
\end{equation}
This matter Lagrangian  provides a natural extension of the $n=1$ effective
matter Lagrangian. 

The kinetic part in the dilaton fields could be diagonalized, however,
we find it more convenient to leave it in the form above which reveals
the symmetry $\Psi_k, \Phi_k \leftrightarrow \Psi_{k'}, \Phi_{k'}$.
The $n$-dimensional cosmological constant leads, through dimensional
reduction, to a Liouville-type potential in the dilaton with 
a coupling constant
$\tilde \Lambda$ related to the $n$-dimensional coupling constant
by means of $ \Lambda_{(4+n)} = 2\alpha^2 \tilde \Lambda $.
The dilatons in this theory thus become massive.

The Lagrangian (\ref{effeL}) is then coupled minimally to Einstein gravity
according to the full action
\begin{eqnarray}
\label{action4}
              S &=& S_G + S_M \nonumber \\
              &=& \int \sqrt{-g^{(4)}} \left(L_G + L_M \right)  d^4 x
\end{eqnarray}
where $L_G = R/(16 \pi G_4)$, $R$ is the Ricci scalar
and $G_4$ is the $4$-dimensional Newton's constant.

Note that the dilaton fields are coupled by an independent coupling
constant $\omega$ to the gauge and Higgs fields. In this respect, the
dilatons here are treated as independent scalar fields, while in
the action (\ref{action}) they appear as parts of the metric tensor.
 
After the rescaling
\begin{equation}
          \Psi_k  =  v \psi_k \ \ , \ \ \omega = \frac{\gamma}{v} \ \ , \ \
\alpha=v\sqrt{G_4}
\end{equation}
the resulting set of equations only depends on the coupling constants
$\alpha$ and $\gamma$. We refrain from giving the explicit form of the
equations here, but refer the reader to \cite{bbh1} for the case $n=1$.

Note that the equations (\ref{mu}), (\ref{A}), (\ref{zeta}), 
(\ref{k}), (\ref{h})
become equivalent to the field equations associated to 
(\ref{action4})
by using the same Ans\"atze for the $4$-dimensional metric, the gauge and
Higgs fields, but by identifying 
\begin{equation}
\zeta_k=2\gamma \psi_k=2\omega \Psi_k \ \ , \ \ \alpha^2=3\gamma^2  \ . 
\end{equation} 
Remarkably, this identification turns out to be independent on $n$.

Note also that the two models differ for $n >1$ with respect to the
constraint (\ref{constraint}) that doesn't exist in the 4-dimensional
effective model.

%%%%%%%%%%%%%%%%%%%%%%%%%%%%%%%%%%%%%%%%%%%%%%%%%%%%%%%%%%%%%%%
\subsection{Non-abelian solutions with diverging Higgs fields}
%%%%%%%%%%%%%%%%%%%%%%%%%%%%%%%%%%%%%%%%%%%%%%%%%%%%%%%%%%%%%%%
So-called $AdS_{n+2}\times S^2$ solutions were first discussed
for $n=1$ and $\Lambda=0$ in \cite{volkov}. They can be thought of
as hypertubes with spherically symmetric cross-section. The generalisation
of these solutions is not possible for $n>1$, since the Higgs
fields don't fulfill the constraints (\ref{constraint}).

However, we can construct the counterparts of these solutions
in the 4-dimensional effective action, since the
constraints don't appear here.

First, we assume  that $H_1(x)=H_2(x)=...\equiv H(x)$ and thus
$\zeta_1(x)=\zeta_2(x)=...\equiv \zeta(x)$.
Further, we choose $\alpha^2=3\gamma^2$ to ``mimick'' the $4+n$-dimensional
case as best as possible.

The fields have the form
\begin{equation}
N(x) = N_0 \ \ , \ \  A(x) = x^{a} \ \ , \ \
\psi_1(x) = \psi_2(x) = \dots = \psi(x) = \frac{\sqrt{3}}{n\alpha} 
\ln\left(\frac{x}{R}\right)
\end{equation}
\begin{equation}
K = \sqrt{q} \ \ , \ \
H_1(x) = H_2(x) = \dots = H(x) = h \left(\frac{x}{R}\right)^{2/n}
\end{equation}
The $00$-component of the energy-momentum tensor for this soltuion reads:
\begin{equation}
T_0^0=-\left(\frac{2N_0h^2}{n}+nqh^2+\frac{(q-1)^2}{2R^4}\right) \ .
\end{equation}
This is constant, however, since $\sqrt{-g}=R^2 x^a \sin\theta$, the energy
of this solution is not finite. Furthermore, the definition
of a magnetic charge cannot be done in the usual sense since
the Higgs field doesn't tend to its vacuum expectation value
asymptotically.

Substituting this ansatz into the equations leads to a set
of five algebraic equations which allow, in principle, to
 determine the parameters
$N_0, a , R , h, q$.
The following relations are easy to obtain:
\begin{equation}
a = \frac{2 + n +2 \kappa h^2}{n}   \ \ , \ \
h^2 = \frac{1-q}{n R^2} \ \ , \ \
N_0 = \frac{q R^2 n^3}{2 n^2 R^2 + 2 \kappa(1-q)} \ \ {\rm with} \ \ 
\kappa\equiv 2\alpha^2
\end{equation}
and leave us with a system of equations for $R$ and $q$.
It turns out that the solution is relatively simple
in the case $n=2$, however much more involved for $n\neq 2$.
This is surely related to the fact that the Higgs field increases
linearly for $n=2$. 

\begin{figure}
\centering
\epsfysize=20cm
\mbox{\epsffile{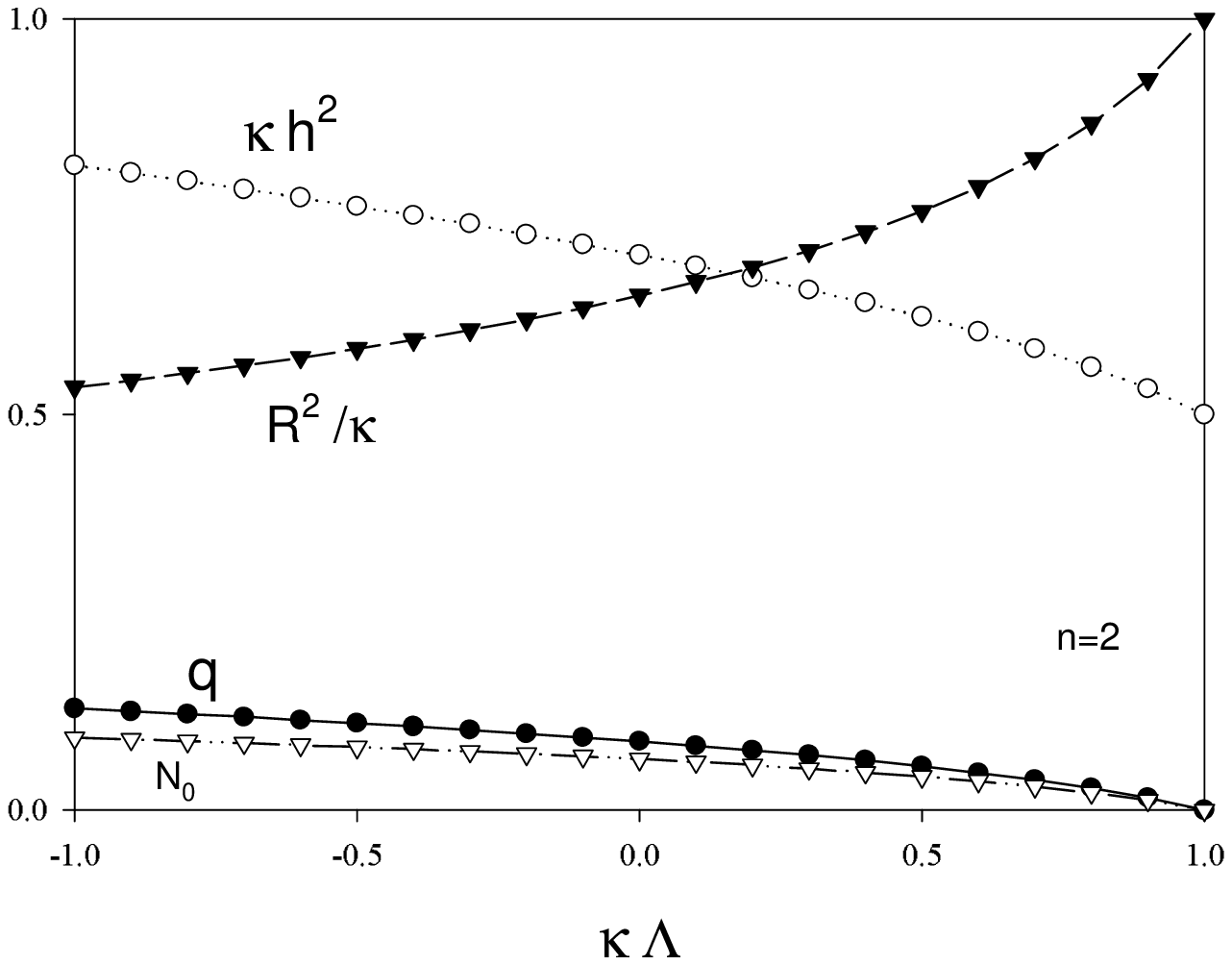}}
\caption{\label{fig1} The values of the constants of the non-abelian
solutions of the 4-dimensional effective model (resulting
for the $4+2$-dimensional model)
are shown in dependence on the cosmological constant.}
\end{figure}

For $n=2$ we find
\begin{equation}
R^2 = \kappa \frac{ 4(1-q)(1-9q^2)- \tilde \Lambda (1-q)^3}
{ 48 q(1+3q)   + \tilde \Lambda (1-q)(3-8q)   } 
\ \ {\rm with} \ \ \tilde{\Lambda}=\kappa \Lambda \ ,
\end{equation}
and the relation determining $q$ is given by the polynomial
equation
\begin{equation}
  \tilde \Lambda^2 (1-q)^4 
  + \tilde \Lambda (1-q)(17-47q+4q^2-24q^3)
  + 16(1+3q)(1-12q+6q^2-9q^3)=0  \ .
\end{equation}
We solved this equation for $q$ numerically by choosing 
 several
values of $\tilde \Lambda$ in $[-1, + 1]$. It turns out that for the values of
$\tilde\Lambda$ considered, only one positive solution
of $q$ exists. Then the various parameters
characterizing the solutions can be computed.
They
are presented in Fig.\ref{fig1} as functions of $\tilde \Lambda$.
As can be seen, both $q$ and $N_0$ tend to zero 
for $\Lambda \rightarrow \frac{1}{2\alpha^2}$.
This tells us that the solutions don't exist for arbitrary values
of the cosmological constant. Specifically, they exist for all values of the
cosmological constant as long as it is chosen to be negative (which corresponds
to $(4+n)$-dimensional Anti-de Sitter space), but for positive cosmological
constant only under the restriction that 
$\Lambda_{(4+n)} < \frac{e^2}{2 G_{(4+n)}}$. 

For generic values of $n$, the solution  is more envolved. We find~:
 \begin{equation}
 R^2 =  \kappa \frac{P}{Q}
 \end{equation}
with
\begin{equation}
P = n^2(q-1)(2n^2q - 2nq^2 + 5nq - n - 2n^2 + 6n)
  ( 2nq + n + 2q)
+\tilde \Lambda (n^3 - 2n^2q - 2nq + 2q)(2q-n) (q-1)^2
 \end{equation} 
and
 \begin{eqnarray}
Q&=&n^2(2n^4 q - 2n^3q^2 + 5n^3q - n^3 - 6n^2q^2 + 6n^2q
- 4nq^2 + 2nq - 4q)(2nq + n + 2q) \nonumber \\
&+&
\tilde \Lambda n(q-1)(- n^5 + 4n^4q - 4n^3q^2 + 4n^3q - 8n^2q^2 - 4n^2
q - 2nq + 8q^2 + 2q) \ .
 \end{eqnarray} 
 The equation for $q$  is a lengthly
polynomial expression of degree nine in
$q$ and of degree three in $\tilde\Lambda$, that's why we don't give it here.

The solution presented above is completely symmetric under the
exchange
\begin{equation}
H_j,\zeta_j \leftrightarrow  H_k,\zeta_k \ \ , \ \ \ 
\forall j,k \in \{1, \dots , n \} \ .
\end{equation}
It can  be generalized in the following way~:
 \begin{equation}
N(x) = N_0 \ \ , \ \  A(x) = x^{a} \ \ , \ \
K(x)=\sqrt{q} 
\end{equation}
and no-equal dilaton and Higgs fields
\begin{equation}
\psi_k(x)  = \frac{\sqrt{3}}{n\alpha} \ln\left(\frac{x}{R_k}\right) \ \ , \ \
H_k(x)  = h \left(\frac{x}{R_k}\right)^{2/n} \ \ , \ \ k=1,...,n \ \ ,
\end{equation}
where $R_1, \dots , R_n$ are positive constants which fulfill 
the constraint  $\displaystyle\prod_{k=1}^n R_k = R^n$, where $R$ is the quantity
determined in the symmetric case. This choice of functions
leads to a family of $n-1$ possible solutions.

\section{Conclusions}
Adopting the dimensional reduction scheme based on the fact
that the fields do not depend on the $n$ extra dimensions, we have studied the
classical equations corresponding to the SU(2)
Einstein-Yang-Mills model
in $4+n$ dimensions including a cosmological constant. 

Supplementing the initial model with a
cosmological constant results in a self-interacting potential of the
dilatons, where this potential is completely fixed in terms of the
coupling constants of the model.

A spherically symmetric Ansatz for the fields leads
to a system of non-linear differential equations. 
Constraints on the fields lead to the fact that only solutions
which become effectively 5-dimensional (with one non-zero Higgs fields)\
or with all Higgs fields constant and zero gauge fields exist. We construct the ``embedded'' abelian
solutions with all Higgs fields constant. The Ricci and Kretschman scalars of these solutions
are finite everywhere apart
from the origin. Depending on the choice of parameters they represent
either naked singularities or black holes.
They generalize
solutions obtained in \cite{chm} to the case of a $(4+n)$-dimensional
space-time with an $SO(3)$ symmetry. These solutions are not asymtotically
flat, neither de Sitter nor Anti-de Sitter,
but all components of the energy-momentum tensor
remain finite asymptotically.
  
One interesting generalisation of this result would be to construct the
rotating counterparts of these solutions, following \cite{gm} for
the Einstein-Maxwell case.

The second family of solutions obtained
in this paper is the family of solutions of the
effective 4-dimensional action. 
These solutions are characterized by non-trivial matter 
fields, especially diverging Higgs fields. 
They depend smoothly on the
cosmological constant.
We find that for $n=2$ these type of solutions exist only up to a maximal
value of the cosmological constant.\\
\\
\\
{\bf Acknowledgements}
Y. B. gratefully acknowledges the Belgian F.N.R.S. for financial support.

 \newpage

 \newpage


\begin{thebibliography}{99}
\bibitem{kaluza}
T. Kaluza, Sitzungsber. Preuss. Akad. Wiss. Berlin (1921).
\bibitem{klein}
O. Klein, Z. Phys. {\bf 37} (1926), 895.
\bibitem{maeda}
G. Gibbons and K. Maeda, Nucl. Phys. {\bf B298} (1988), 741;
D. Garfinkle, G. Horowitz and A. Strominger, Phys. Rev. {\bf D43} (1991), 371.
\bibitem{pol} see e.g. J. Polchinski, {\it String Theory}, Cambridge University press (1998).
\bibitem{brane} K. Akama, ``Pregeometry'',
in Lecture Notes in Physics, 176, Gauge Theory
and Gravitation, Proceedings, Nara, 1982, edited by K. Kikkawa, N. Nakanishi
and H. Nariai, 267-271 (Springer-Verlag, 1983) [hep-th/0001113];
V. A. Rubakov and M. E. Shaposhnikov, Phys. Lett. {\bf 125 B} (1983), 136; {\bf 125 B} (1983), 139;
G. Davli and M. Shifman, Phys. Lett. {\bf B 396} (1997), 64; {\bf 407} (1997), 452;
I. Antoniadis, Phys. Lett. {\bf B 246} (1990), 377; 
N. Arkani-Hamed, S. Dimopoulos and G. Dvali, Phys. Lett. {\bf B 429} (1998), 263;
I. Antoniadis, N. Arkani-Hamed, S. Dimopoulos and G. Dvali, Phys. Lett. 
{\bf B 436} (1998), 257; L. Randall and R. Sundrum, Phys. Rev. Lett. {\bf 83} (1999), 3370; {\bf 83} (1999), 4690. 

\bibitem{ads} J. Maldacena, Adv. Theor. Phys. {\bf 2} (1998), 231;
E. Witten, Adv. Theor. Math. Phys. {\bf 2} (1998), 253;
{\it for a review see e.g.} O. Aharony, S. Gubser, J. Maldacena, H.
Ooguri and Y. Oz, Phys. Rep. {\bf 323} (2000), 183.
\bibitem{ds} A. Strominger, JHEP {\bf 10} (2001), 034; JHEP {\bf 11} (2001), 049.
\bibitem{bcht} Y. Brihaye, A. Chakrabarti, B. Hartmann and D. H. Tchrakian,
Phys. Lett. {\bf B 561} (2003), 161.

\bibitem{volkov}
M. S. Volkov, Phys. Lett. {\bf B524} (2002), 369.
\bibitem{bch} Y. Brihaye, F. Clement and B. Hartmann, {\it Spherically symmetric
Yang-Mills solutions in a ($4+n$)-dimensional space-time}, hep-th/0403041.
\bibitem{bbh1} B. Hartmann, Y. Brihaye and B. Bertrand, Phys. Lett. {\bf B 570} (2003), 137.
\bibitem{pw} S. Poletti and D. Wiltshire, Phys. Rev. {\bf D 50} (1994), 7260.
\bibitem{chm} K. C. K. Chan, J. H. Horne and R. B. Mann, 
Nucl. Phys. {\bf B447} (1995), 441.
\bibitem{weinberg}
K. Lee, V.P. Nair and E.J. Weinberg,
 Phys. Rev. {\bf D45} (1992), 2751;\\
 P. Breitenlohner, P. Forgacs and D. Maison,
 Nucl. Phys. {\bf B383} (1992), 357;\\
 P. Breitenlohner, P. Forgacs and D. Maison,
 Nucl. Phys. {\bf B442} (1995), 126.
\bibitem{thooft}
 G. `t Hooft, 
%Magnetic monopoles in unified gauge theories, 
 Nucl.~Phys.~ {\bf B79} (1974), 276;\\
 A.~M. Polyakov, 
%Particle spectrum in quantum field theory,
 JETP Lett. {\bf 20} (1974), 194.

\bibitem{wiltshire} D. Wiltshire, Phys. Rev. {\bf D 44} (1991), 1100.
\bibitem{gm} T. Ghosh and P. Mitra, Class. Quantum Grav. {\bf 20} (2003), 1403.
\end{thebibliography}
\end{document}